\documentclass[twocolumn,aps,pre,showpacs,amsfonts,superscriptaddress,byrevtex]{revtex4}
\begin{document}
%----------------------------------------------------------------------------
\title{Galilean invariance and homogeneous anisotropic randomly stirred
flows
}
%--------------------------------------------------------------------------
\author{Arjun Berera}
\email{ab@ph.ed.ac.uk}  \affiliation{School of Physics, University
of Edinburgh, Edinburgh, EH9 3JZ, U.K.}
\author{David Hochberg}
\email{hochberg@laeff.esa.es} \affiliation{Centro de
Astrobiolog\'{\i}a (CSIC-INTA), Ctra. Ajalvir Km. 4, 28850
Torrej\'{o}n de Ardoz, Madrid, Spain}
%--------------------------------------------------------------------------

\begin{abstract}
The Ward-Takahashi (WT) identities for incompressible flow implied
by Galilean invariance are derived for the randomly forced
Navier-Stokes equation (NSE), in which both the mean and
fluctuating velocity components are explicitly present. The
consequences of Galilean invariance for the vertex renormalization
are drawn from this identity.
\end{abstract}

\date{\today}

\maketitle

The transformation properties of the Navier-Stokes equation (NSE)
play an important part in both the physics of turbulence as well
as in practical aspects of turbulence modelling \cite{Pope}. Among
the invariance properties or symmetries of the NSE, Galilean
invariance is perhaps the most frequently cited in theoretical
approaches to turbulence \cite{FNS,DM,T,CT,Mou}.

For fluctuating flows, a decomposition of the instantaneous fluid
velocity into a mean plus a component fluctuating relative to the
mean (the Reynolds decomposition) shows that the mean and
fluctuating velocities respond to a Galilean transformation in
different ways. This would seem to be the most general physical
significance one could deduce by studying the behavior of the NSE
under Galilean transformations. On the other hand, because the
Galilean transformation is a symmetry of the NSE, it leads to
nontrivial and exact identities among various correlation
functions. These so-called Ward-Takahashi (WT) identities make two
types of important statements: ($a$) In quantum field theory
(QFT), they relate formally divergent parts of different
correlation functions. Ward-Takahashi identities reduce the number
of counterterms or renormalization constants; and ($b$) in QFT as
well as in field theoretic approaches to condensed matter
problems, they relate different physical quantities. For example,
in hydrodynamics they relate the eddy viscosity and the numerical
prefactor multiplying the nonlinear convective term (the vertex).
Thus, the deeper physical significance of the Galilean
transformation is that it \textit{constrains} the values of the
physical parameters appearing in the NSE, and in particular, it
leads to the oft cited non-renormalization of the vertex of the
NSE in the \textit{large-scale flow} regime \cite{FNS}.

Recently however it was claimed that Galilean invariance does not
at all constrain the vertex renormalization \cite{Mc}, thus
putting into serious question the widely cited key early papers in
turbulence research \cite{FNS,DM}. In \cite{Mc}, a distinction is
drawn between the NSE for the instantaneous velocity field $V_i$
and the NSE for the fluctuating velocity $u_i$ in the comoving
frame of the mean constant velocity, for spatially homogeneous
flow. These velocities are defined through the Reynolds
decomposition of the velocity into its mean plus fluctuations:
$V_i(\mathbf{x},t) = \langle V_i(\mathbf{x},t)\rangle +
u_i(\mathbf{x},t)$, where the angular brackets denote the ensemble
average \cite{Pope}. The essential observation made in \cite{Mc}
is that Galilean invariance of the equation for the fluctuating
velocity $u_i$ is established by transforming the mean velocity
only. No previous field theoretic derivations of the vertex
non-renormalization result have distinguished between the mean and
fluctuating velocity components, since they treat the
instantaneous velocity field at zero mean velocity, in other
words, the case of a homogeneous and isotropic flow. In order to
address specifically the concern in \cite{Mc}, in which the mean
velocity plays a distinguished role, one should examine carefully
the derivation and physical consequence of the pertinent WT
identities, keeping explicit the decomposition of the
instantaneous velocity field into its fluctuating and generally
nonzero mean components. In short, one needs to treat the case of
a homogeneous but anisotropic fluid. The purpose of this Brief
Report is to carry this out in some detail, paying careful
attention to the rather different ways the mean and fluctuating
velocities are affected by a Galilean transformation. The issue of
vertex renormalization will be addressed within this framework.

We start with the Reynolds decomposition to express the NSE in
terms of the constant mean velocity and the fluctuation about the
mean. So we take $V_i = K_i + u_i$, where $K_i= \langle V_i
\rangle$ is the constant mean velocity and $u_i$ is the
fluctuation about the mean. By definition, $\langle u_i \rangle =
0$. As demonstrated in \cite{Mc}, the equation of motion for a
constant mean velocity $K_i$ vanishes identically, resulting in
the NSE for the fluctuating velocity $u_i$. To this we add a
stirring force $f_i$ and express the equation in solenoidal form:
\begin{equation}\label{fNSEu}
\frac{\partial u_i}{\partial t} + K_j \frac{\partial u_i}{\partial
x_j}  + \lambda_0 P_{ij}(\nabla ) \frac{\partial (u_l
u_j)}{\partial x_l} = \nu_0 \nabla^2 u_i + f_i,
\end{equation}
where the projection operator is $P_{ij}(\nabla ) =
\big(\delta_{ij} - \nabla_i \frac{1}{\nabla^2} \nabla_j \big)$.
$f_i$ is Gaussian with zero mean $\langle f_i \rangle = 0$ and
time and space translation invariant: $\langle f_i({\bf
x},t)f_j({\bf y},\tau)\rangle = D_{ij}({\bf x} - {\bf y})\delta(t
- \tau)$. The viscosity is written with subscript zero, $\nu_0$,
to remind us that this is the bare (unrenormalized) molecular
viscosity that appears in the fluid equation of motion. We provide
the nonlinear term in Eq. (\ref{fNSEu}) with a parameter
$\lambda_0 = 1$. This is a well-known convenient bookkeeping
device used in perturbation theory, in which typically, solutions
of Eq. (\ref{fNSEu}) are developed formally in powers of
$\lambda_0$.

As we are interested in the correlation functions implied by Eq.
(\ref{fNSEu}), we make use of dynamic functional methods
\cite{DeDom,Jan,Phythian,Jensen}. The generating functional for
the correlation functions associated with Eq. (\ref{fNSEu}) is
given by
\begin{eqnarray}\label{Zu}
Z[{\bf J}, {\bf \Sigma}]_{K} &=& \int [\mathcal{D}{\bf
u}][\mathcal{D} {\bf \sigma}]
\exp \left\{ -S[{\bf u}, {\bf \sigma}]_{K} \right. \nonumber \\
&+& \left. \int d{\bf x}\, dt ({\bf J \cdot u} + {\bf J \cdot K}+
{\bf \Sigma \cdot \sigma}) \right\},
\end{eqnarray}
where the action $S$ is
\begin{eqnarray}\label{classu}
&&S[{\bf u}, {\bf \sigma}]_{K} = \int d{\bf x} dt \left(
\textstyle{\frac{1}{2}}\int d{\bf y} \,{\bf \sigma}_i({\bf
x},t)D_{ij}({\bf x} - {\bf y}) {\bf
\sigma}_j({\bf y},t) \right. \nonumber \\
&& -i\left. \sigma_k \Big(\frac{\partial u_k}{\partial t} + K_j
\frac{\partial u_k}{\partial x_j}  + \lambda_0 P_{kj}(\nabla )
\frac{\partial (u_l u_j)}{\partial x_l} - \nu_0 \nabla^2 u_k \Big)
\right).\nonumber\\
\end{eqnarray}
The field $\bf {\sigma}$ is conjugate to the velocity $u_i$, and
is the analog of the noncommuting operator introduced by Martin,
Siggia and Rose (MSR) in their operator formalism of classical
statistical dynamics \cite{MSR}. $\bf J$ and $\bf \Sigma$ are
arbitrary space and time dependent source functions for $\bf V = u
+ K$ and $\bf \sigma$, respectively. It is important to note that
both the mean and fluctuating velocities couple to the same
external source. A straightforward and convincing way to see that
this must be so is to start from the forced NSE for the
instantaneous velocity $V_i$ and set up its corresponding action
and dynamic functional. Couple $V_i$ to a source term $J_i$.
Insert the Reynolds decomposition $V_i = K_i + u_i$ into the
measure, action and into the source terms. The incompressibility
of the flow implies the fluctuations are themselves
incompressible: $\nabla_i V_i = 0 \Rightarrow \nabla_i u_i = 0$.
Then Eqs. (\ref{Zu}) and (\ref{classu}) follow immediately from
the dynamic functional and action for the instantaneous velocity.

We now subject the generating functional $Z$ to a Galilean
transformation. Consider a second primed frame moving with
constant velocity $\mathbf{c}$ with respect to the unprimed frame.
Then the relations between time and coordinates of an event in the
unprimed and primed frames are $ t = t'$ and ${\bf x} = {\bf c}t +
{\bf x}'$ while the velocities and conjugate field transform as
$K_i = c_i + K'_i,\, u_i({\bf x}, t) = u'_i({\bf x'}, t')$ and
${\sigma_i({\bf x}, t)} = {\sigma'_i({\bf x'}, t')}.$ Note that
the mean and fluctuating velocities are affected in different
ways. The conjugate field $\sigma$ has dimensions of acceleration
and so transforms as indicated above. Using the time and
coordinate transformations as well as these rules, it is easy to
verify that $S[{\bf u'}, {\bf \sigma'}]_{K'} = S[{\bf u}, {\bf
\sigma}]_K$ and the functional integral measures are invariant:
$[\mathcal{D}{\bf u'}]= [\mathcal{D}{\bf u}]$, $[\mathcal{D} {\bf
\sigma'}]= [\mathcal{D} {\bf \sigma}]$.

To obtain the Ward-Takahashi identity, consider an
\textit{infinitesimal} Galilean transformation and Taylor expand
the fluctuating velocity out to first order in the relative
velocity $\delta {\bf c}$ as follows:
\begin{eqnarray}\label{tesimal}
{\bf u}( {\bf x} + \delta {\bf c}\,t, t) &=& {\bf u}( {\bf x}, t)
+ t(\delta {\bf c}\cdot \nabla){\bf u}( {\bf x}, t) ,
\end{eqnarray}
with an analogous expansion for $\sigma$. This implies that the
functional $Z$ in Eq. (\ref{Zu}) transforms as $Z \rightarrow Z +
\delta Z$, where
\begin{equation}\label{deltaZ}
\delta Z = \big\langle \Big( \int t J_l (\delta {\bf c} \cdot
\nabla) u_l + t \Sigma_l (\delta {\bf c} \cdot \nabla) \sigma_l -
{\bf J \cdot}\delta {\bf c} \Big) \big\rangle_{\bullet}
 = 0.
\end{equation}
The angular brackets with bullet denote the average taken over the
turbulent ensemble in \textit{the presence of the source terms}.
As Eq. (\ref{deltaZ}) holds for all $\delta {\bf c}$, it follows
that $Z$ satisfies the functional differential equation (the
integral is over $\mathbf{x},t$)
\begin{equation}
\label{WTZu} \Big( \int t J_l \nabla_j \frac{\delta}{\delta J_l} +
t \Sigma_l \nabla_j \frac{\delta}{\delta \Sigma_l}- J_j\Big)
Z[{\bf J},{\bf \Sigma}]_{K} = 0.
\end{equation}
This makes use of  the fact that the mean velocity $K_l$ is
constant in homogeneous turbulence, hence $\nabla_j \langle K_l
\rangle_{\bullet} = \nabla_j K_l \langle 1 \rangle_{\bullet} = K_l
\nabla_j\langle 1 \rangle_{\bullet} = 0$, and so
$\nabla_j\frac{\delta}{\delta J_l}Z[{\bf J},{\bf \Sigma}]_{K} =
\nabla_j \langle u_l + K_l \rangle_{\bullet} = \nabla_j \langle
u_l \rangle_{\bullet} + \nabla_j \langle K_l \rangle_{\bullet} =
\nabla_j \langle u_l \rangle_{\bullet}.$ This identity Eq.
(\ref{WTZu}) for the functional $Z$ can be used to obtain exact
relationships among the correlation functions associated with the
random velocity field in its Reynolds decomposed form. Of course,
$Z$ generates both the disconnected and connected correlation
functions. For the purposes of renormalization, it is customary to
work instead with the connected proper vertices, also known as the
one-particle irreducible (1PI) Green functions \cite{ZJ}. It is
important to point out that at lowest order, or tree-level, the
connected proper vertex functions coincide with the vertices of
the original (bare) action $S$ (\ref{classu}). To obtain the
generating functional for 1PI vertices, we follow standard
practice and introduce the generating functional of the connected
Green functions $W = \ln Z$, and then carry out a Legendre
transform \cite{ZJ} $\Gamma[\langle {\bf V} \rangle_{\bullet},
\langle {\bf \sigma} \rangle_{\bullet}] = -W[{\bf J}, {\bf
\Sigma}] + \int {\bf J \cdot} \langle {\bf V} \rangle_{\bullet} +
{\bf \Sigma \cdot} \langle {\bf \sigma} \rangle_{\bullet}$, where
\begin{equation}\label{subsid}
\langle V_k \rangle_{\bullet} = \frac{\delta W}{\delta J_k},\,
\langle \sigma_k \rangle_{\bullet} = \frac{\delta W}{\delta
\Sigma_k},\, \frac{\delta \Gamma}{\delta \langle V_k
\rangle_{\bullet}} = J_k,\, \frac{\delta \Gamma}{\delta \langle
{\sigma}_k \rangle_{\bullet}} = \Sigma_k.
\end{equation}
Note of course that $\langle V_k \rangle_{\bullet} = \langle u_k
\rangle_{\bullet} + K_k $. From the definitions of $W$ and
$\Gamma$ and using Eq. (\ref{subsid}), we obtain immediately the
Ward-Takahashi identity satisfied by $\Gamma$:
\begin{equation}\label{WTG}
\int \Big( t \nabla_i \langle {V}_l \rangle_{\bullet} \frac{\delta
\Gamma}{\delta \langle V_l \rangle}_{\bullet} + t \nabla_i \langle
{\sigma}_l \rangle_{\bullet} \frac{\delta \Gamma}{\delta \langle
{\sigma}_l \rangle}_{\bullet} - \frac{\delta \Gamma}{\delta
\langle {V}_i \rangle}_{\bullet} \Big) = 0.
\end{equation}
The great utility of the WT identity Eq. (\ref{WTG}) is that it
provides \emph{exact}, non-perturbative equations relating various
proper vertices (correlation functions) associated with the NSE
equation. At lowest order (i.e., tree level), these proper
vertices can be read off directly from the action $S$ (see below).
Here, and in conformity with the Reynolds decomposition, the mean
value of the velocity field in the \textit{absence} of source
terms is a constant (and nonzero) vector: $\langle V_i ({\bf x},
t) \rangle = \delta W[{\bf J},{\bf \Sigma}]/{\delta
J_i}|_{J=\Sigma = 0} = K_i.$ A nonzero expectation value for the
fluid velocity in the limit of vanishing external sources is
reminiscent of symmetry breaking in field theory. Indeed, the
existence of a constant $K_i$ breaks the fluid isotropy since a
preferred direction is being singled out. In this case, the
correct Taylor series representation of the effective action
$\Gamma$ is given by \cite{AbersLee}
\begin{eqnarray}\label{Taylor2}
&{}&\,\,\Gamma[\langle{\bf V}\rangle_{\bullet}, \langle{\bf
\sigma}\rangle_{\bullet}] = \sum_{m1,m2 = 1}^{\infty} \frac{1}{m1!
m2!} \int [{d\bf x} \, dt] \nonumber \\
&{}&\Gamma^{(m1,m2)}_{i1,i2,...,im1,j1,j2,...,jm2} \big(\langle
V_{i1}(1)\rangle_{\bullet}- K_{i1}\big)...\times \nonumber\\
&{}&\big(\langle V_{im1}(m1)\rangle_{\bullet}- K_{im1}\big)
\langle\sigma_{j1}(1)\rangle_{\bullet} ...\langle
\sigma_{jm2}(m2)\rangle_{\bullet}
\end{eqnarray}
where
\begin{widetext}
\begin{equation}\label{coefficient}
\Gamma^{(m1,m2)}_{i1,i2,...,im1,j1,j2,...,jm2} = \left.
\frac{\delta^{m1 + m2} \Gamma[\langle{\bf V}\rangle_{\bullet},
\langle{\bf \sigma}\rangle_{\bullet}]}{\delta \langle V_{i1}(1)
\rangle_{\bullet} ...\delta \langle V_{im1}(m1) \rangle_{\bullet}
\delta \langle \sigma_{j1}(1) \rangle_{\bullet}...\delta \langle
\sigma_{jm2}(m2) \rangle_{\bullet}}\right|_{\stackrel{\langle V
\rangle_{\bullet} = K}{\langle \sigma \rangle_{\bullet} = 0}}.
\end{equation}
\end{widetext}
We employ a condensed notation to indicate the dependence on space
and time, thus, e.g., $(1) \equiv ({\bf x}_1,t_1)$, $(m1) \equiv
({\bf x}_{m1},t_{m1})$ and $[{d\bf x} \, dt]$ stands for the
volume element for all $m1 + m2$ pairs of space and time points to
be integrated over in each term of the sum in Eq. (\ref{Taylor2}).
The factors Eq. (\ref{coefficient}) for each $m1$ and $m2$,
correspond to specific proper vertex with $m1$ factors of the
velocity and $m2$ factors of the conjugate field. So, for example,
the inverse response function and proper vertex associated with
the nonlinear convective term in the NSE are (suppressing the
dependence on the space and time arguments) $\Gamma^{(1,1)}_{ij}$,
and $\Gamma^{(2,1)}_{ijk}$, respectively.

The pertinent WT identity that we seek is obtained by inserting
Eq. (\ref{Taylor2}) into Eq. (\ref{WTG}) differentiating the
latter with respect to $\frac{\delta }{\delta  \langle V_k({\bf
y}, t')\rangle_{\bullet}} \frac{\delta }{\delta \langle
\sigma_j({\bf w}, t'')\rangle_{\bullet}}$, using the definition in
Eq. (\ref{coefficient}) and then setting $\langle V
\rangle_{\bullet} = K$ and  $\langle \sigma \rangle_{\bullet} =
0$. Doing so yields
\begin{eqnarray}\label{identity1}
(t'' &-& t')\frac{\partial }{\partial y_i}\Gamma_{jk}^{(1,1)}({\bf
y}, t',{\bf w}, t'')\nonumber \\
&=& \int d{\bf x}\,dt \, \Gamma^{(2,1)}_{ijk} ({\bf x}, t, {\bf
y}, t', {\bf w}, t''),
\end{eqnarray}
which follows after an integration by parts (we adopt vanishing
boundary conditions) and using the fact that spatial homogeneity
implies proper vertices are translational invariant in
configuration space. For our final step, we express this identity
in terms of wavenumber (or, momentum) and frequency space $({\bf
k}, \omega)$, by means of the Fourier transform (FT). Recall that
for stationary flow, the proper vertices can depend only on time
\textit{differences}. Thus, the FT of Eq. (\ref{identity1}) yields
\begin{equation}\label{identity3b}
-k_i \frac{\partial}{\partial \omega} \Gamma^{(1,1)}_{jk}({\bf
k},\omega; -{\bf k},-\omega) = \Gamma_{ijk}^{(2,1)}({\bf 0}, 0;
{\bf k},\omega; -{\bf k},-\omega).
\end{equation}
At zero-loop order (indicated via the zero superscript), and from
Eq. (\ref{classu}) we can read off directly the 1PI functions that
are related by this WT identity, namely
\begin{eqnarray}\label{bare1}
\stackrel{0}{\Gamma}^{(1,1)}_{ij} &=& \Big(-i\omega + iK_j k_j + \nu_0 k^2 \Big)P_{ij}({\bf k}) \\
\label{bare2} \stackrel{0}{\Gamma}^{(2,1)}_{ijk} &=& i\lambda_0 \,
k_i P_{jk}({\bf k}).
\end{eqnarray}
It is important to realize that the $K_j$-dependence of the bare
response function (\ref{bare1}) follows from applying the Reynolds
decomposition from the outset to the NSE, which leads to
(\ref{fNSEu}) and (\ref{classu}). This expression \emph{cannot} be
obtained from applying a Galilean transformation to the response
function for an isotropic fluid ($K_j=0$). Inserting Eqs.
(\ref{bare1},\ref{bare2}) into Eq. (\ref{identity3b}) implies that
$\lambda_0 = 1$. In other words, Galilean invariance requires that
the bookkeeping parameter introduced in Eq. (\ref{fNSEu}) must be
identically unity. This result is clearly independent of the mean
velocity $K_i$.  To go further, we make the reasonable assumption
that some of the perturbative corrections to the solution of Eq.
(\ref{fNSEu}) can be absorbed into redefinitions of the parameters
appearing in the equation of motion. This is to say, we assume the
NSE is partially \textit{form-invariant} with respect to the
renormalization arising from the fluctuations inherent in the
randomly stirred ensemble. This renormalization leads to frequency
and wavevector dependent viscosity and a ``mass" term $\Sigma$ in
the inverse response function. The vertex associated with the
convective term in the NSE can also in principle be modified or
suffer renormalization, leading to a frequency and wavevector
dependent term which we denote $\Lambda$. The most general terms
one can write down which maintain the same \emph{tensorial
structure} as their unrenormalized counterparts Eq. (\ref{bare1})
and Eq. (\ref{bare2}) are
\begin{eqnarray}\label{warda}
&&\Gamma^{(1,1)}_{jk}({\bf k},\omega; -{\bf k},-\omega) =
\nonumber \\
&&\qquad \Big( -i\omega + \nu(\omega,k) k^2 + \Sigma(\omega,k ) \Big) P_{jk}({\bf k}) \\
\label{wardb} &&\Gamma_{ijk}^{(2,1)}({\bf 0}, 0; {\bf k},\omega;
-{\bf k},-\omega) = i k_i \,P_{jk}({\bf k})+
\Lambda_{ijk}(\omega,{\bf k}),\nonumber \\
&{}&
\end{eqnarray}
where in accordance with Eq. (\ref{identity3b}) and Eqs.
(\ref{bare1},\ref{bare2}), we have set $\lambda_0 = 1$ in Eq.
(\ref{wardb}). The relation of Eq. (\ref{warda}) and Eq.
(\ref{wardb}) to Eq. (\ref{bare1}) and Eq. (\ref{bare2}),
respectively, is as follows. Imagine setting the nonlinear term in
the NSE to zero by \textit{formally} taking $\lambda_0 \rightarrow
0$. Then $\nu(\omega,k) \rightarrow \nu_0$, $\Sigma(\omega,k)
\rightarrow iK_jk_j$ and $\Lambda_{ijk}(\omega,k) \rightarrow 0$.
That is, these terms appear as a result of the perturbation
expansion, which is developed in powers of $\lambda_0$. Finally,
inserting these expressions Eqs. (\ref{warda},\ref{wardb}) into
the Ward-Takahashi identity Eq. (\ref{identity3b}) implies that
\begin{equation}\label{ward3}
-k_iP_{jk}({\bf k}) \Big( k^2 \frac{\partial
\nu(\omega,k)}{\partial \omega} + \frac{\partial
\Sigma(\omega,k)}{\partial \omega} \Big) =
\Lambda_{ijk}(\omega,{\bf k}).
\end{equation}
There are some general consequences one can draw from this
identity. First of all, in models of stationary forced turbulence
Eq. (\ref{fNSEu}), the effective viscosity can not depend on time
(nor will its FT depend on the frequency $\omega$), and thus at
most the renormalized viscosity could depend on wavenumber: $\nu =
\nu(k^2)$. The same holds for the ``mass" term $\Sigma =
\Sigma(k)$, so that for stationary random forcing Eq.
(\ref{ward3}) immediately implies that $\Lambda_{ijk}(\omega,{\bf
k}) = 0$, and hence the vertex function $\Gamma_{ijk}^{(2,1)}$ in
Eq. (\ref{wardb}) is not altered by perturbative corrections to
the equation of motion. When $\Lambda_{ijk}(\omega,{\bf k}) = 0$,
this vertex is that which corresponds to the convective term in
the NSE Eq. (\ref{fNSEu}). There is one important caveat here:
this conclusion holds only for the case of \emph{zero-momentum
transfer} in the vertex (see Eq. (\ref{identity3b})). We are not
claiming that the vertex cannot have nontrivial renormalization
for \textit{finite} momentum transfer. This result of vanishing
vertex correction $\Lambda_{ijk}(\omega,{\bf k})$ is valid only
for large-scale flows and is \textit{independent} of the value of
$K_i$.

In conclusion, when we come to reconcile Galilean invariance with
fluctuation phenomena, we know that the enforcement of this
symmetry does impose nontrivial conditions on some of the physical
parameters appearing in the equation of motion. For example,
recursive renormalization group (RG) procedures applied to the NSE
explicitly generate higher-order nonlinearities in the
renormalized fluid equation of motion. These higher-order
nonlinearities, such as triple velocity products, do however
preserve Galilean invariance \cite{ZV}. Correlation functions
provide further information that can not be ascertained at the
level of the evolution equation of motion alone. Here it is
worthwhile to recall that the original operator-based perturbation
theory developed by MSR requires three different vertices. They
claim that the whole problem of strong turbulence is contained in
a proper treatment of the vertex renormalization \cite{MSR}. The
two additional vertices introduced by MSR have no counterpart at
the level of the NSE.

In contrast to the NSE, correlation functions can not be neatly
separated into mean and fluctuating parts. This in general means
constraints from Galilean invariance for the mean velocity
component also can have implications for the fluctuating
component. In particular, this leads to the WT identity Eq.
(\ref{ward3}), which provides a nontrivial relation between the
fluid response function and the nonlinear vertex term (convective
term). This dependence holds of course only for the large scale
flow regime and says nothing about the small scale flow behavior
i.e., the inertial and dissipation ranges). This is so because
(\ref{ward3}) follows directly from (\ref{identity3b}), which as
can be checked, involves the vertex only for zero momentum
transfer.

In \cite{Mc}, the claim is made that Galilean invariance provides
no constraint whatsoever on the vertex renormalization, in sharp
contrast to the results in \cite{FNS,DM,CT}. These previous papers
that examined the WT identities did so in the frame at zero mean
velocity in which it would be impossible to test the claims of
\cite{Mc}, which depend crucially on having Reynolds decomposed
the fluid velocity.  To test these claims requires working in a
frame of general nonzero mean velocity. This paper has done this
and demonstrated that although the claims in \cite{Mc} have some
validity, the fluctuating velocity component is still partly
constrained, in particular in the limit of zero momentum transfer.
Thus our result provides a resolution to the seemingly differing
conclusions drawn in \cite{Mc} and in \cite{FNS,DM,CT}. In the
process of addressing this issue, our paper has also provided the
most general form of the WT identities for the randomly forced
NSE. These results can not be derived from the other treatments of
the WT identities in \cite{FNS,DM,CT} by simply performing a
Galilean transformation of their results. Further understanding
about the implications of Galilean invariance in the NSE can
perhaps be gained by examining the analogy between this symmetry
and global gauge invariance in QFT and exploring the implications
of gauge fixing procedures.

We thank R. Horsley and W.D. McComb for helpful discussions. A.B.
was funded by the U.K. Particle Physics and Astronomy Research
Council (PPARC). D.H. acknowledges a U.K. Royal Society funded
Study Visit and the CSIC ``Marina Bueno" program.


\begin{thebibliography}{99}
\bibitem{Pope} S.B. Pope, \textit{Turbulent Flows} (Cambridge
University Press, Cambridge, 2000).
\bibitem{FNS} D. Forster, D.R. Nelson and M.J. Stephen, Phys. Rev.
A{\bf 16}, 732 (1977).
\bibitem{DM} C. DeDominicis and P.C. Martin, Phys. Rev. A{\bf 19},
419 (1979).
\bibitem{T} E.V. Teodorovich, J. Applied Math. and Mech. {\bf 53},
340 (1989).
\bibitem{Mou} C-H. Mou and P.B. Weichmann, Phys. Rev. E{\bf 52},
3738 (1995).
\bibitem{CT} R. Collina and P. Tomassini, hep-th/9709185v2, (1997).
%October 1997.
\bibitem{Mc} W.D. McComb, Phys. Rev. E\textbf{71}, 037301 (2005).
\bibitem{DeDom} C. DeDominicis, J. Phys. (Paris) Colloq. {\bf 1},
247 (1976).
\bibitem{Jan} H.K. Janssen, Z. Phys. B {\bf 23}, 377 (1976).
\bibitem{Phythian} R. Phythian, J. Phys. A {\bf 10}, 777 (1977).
\bibitem{Jensen} R.D. Jensen, J. Stat. Phys. {\bf 25}, 183 (1981).
\bibitem{MSR} P.C. Martin, E.D. Siggia and H.A. Rose, Phys. Rev.
A{\bf 8}, 423 (1973).
\bibitem{ZJ} J. Zinn-Justin, \textit{Quantum Field Theory and
Critical Phenomena} (Oxford University Press, Oxford, 2002) 4rth
ed.
\bibitem{AbersLee} E.S. Abers and B.W. Lee, Phys. Lett. 9{\bf C},
1-141 (1973).
\bibitem{ZV} Y. Zhou and G. Vahala, Phys. Rev. E {\bf 48}, 4387
(1993).
\end{thebibliography}
\end{document}